\documentstyle[twocolumn,aps,psfig,floats]{revtex}
\begin{document}
\draft
\title{An Approximate Sign Sum Rule for the Transmission Amplitude
  Through  
  a Quantum Dot}  
\author{Reinhard Baltin $^1$ and Yuval Gefen $^2$}
\address{$^1$ Max Planck Institut f\"ur Kernphysik, Postfach 103980,
  69029
  Heidelberg, Germany\\
  $^2$ Department of Condensed Matter Physics, The Weizmann Institute
  of Science,
  Rehovot 76100, Israel} 
\date{\today}
\maketitle
\begin{abstract}
    We study the phase of the transmission amplitude through a 
  disordered quantum dot in the Coulomb blockade regime. 
  We calculate the phase dependence on gate voltage for a disorder 
  configuration. We show that over a ``period'', consisting of a 
  resonance and a transmission valley, the total phase change is 0 
  (mod $2\pi$). Deviations from this sum rule are small in the 
  parameter (level spacing/charging energy). The disorder-averaged 
  phase-phase correlation function is found showing 
  interaction-induced correlations between phases at different gate 
  voltages. 
\end{abstract} 
 
The physics of quantum dots (QD) has been the subject of intense 
theoretical and experimental studies in recent years. 
 Most of those studies focused on the role of the 
Coulomb Blockade (CB) in defining quantum transport through such 
systems. Two novel interference experiments \cite{yacoby,schuster} 
studied coherence and transmission phase -- rather than transmission 
probability -- evolution in QD as function of the Aharonov-Bohm flux,
$\Phi$ and the gate voltage, $V_g$. The flux controls the relative
phase through the two arms of the interferometer while the gate
voltage drives the dot in and out of resonance, controlling the mean
number of electrons on the dot. The first experiment \cite{yacoby} -- 
employing a two terminal set-up -- restricted the relative phase of the
interferometer (at $\Phi=0$) to be either $0$ or $\pi$. In the second 
experiment the set-up was similar to a two-slit interferometer and the
(relative) phase of the transmission amplitude through a QD in the CB 
regime could be measured.
 
Several interesting aspects of the experimental data have been 
subsequently discussed in the literature, see 
e.g. \cite{works}. The most
intriguing result though, discussed below, has remained unaccounted
for. It has to do with the evolution of the transmission phase through
the dot, $\alpha$, as $V_g$ is varied, scanning resonances 
and the ``transmission valleys'' between them. Hereafter we attach an 
index $N$ to a valley, corresponding to the number of electrons on the
dot over that range of $V_g$. The resonance separating the valleys
$N-1$ and $N$ will be denoted by $(N-1,N)$. We shall parametrize $V_g$
in the valley by $x$: for the valley $N$ the parameter $x\rightarrow0$
corresponds to the right of resonance $(N-1,N)$, i.e. the point where the
energies of the dot with $N-1$ electrons and $N$ electrons are
practically degenerate; $x\rightarrow1$ corresponds to the left of resonance
$(N,N+1)$. A complete description of $V_g$ is given by the two 
variables $(N,x)$. The remarkable result of the experiment is that as
$V_g$ is varied , the change 
in the transmission phase $\Delta \alpha$ between two
consecutive valleys turns out to be $0$ (mod(2$\pi$)). This is in
distinct contrast to the behavior of non-interacting electrons in
one-dimensional symmetric dots, where $\Delta \alpha=\pi$, or 
two-dimensional
dots where, depending on the details of the geometry and disorder,
$\Delta \alpha$ forms a sequence of $0$ and $\pi$.

There is a large number of works addressing the remarkable
transmission phase correlations observed in the experiments. While we
shall not present here a critical review of all those attempts, it is
worthwhile noting that each of those works is subject to at least one
of the following critiques: (i) An implicit assumption is made
concerning the matrix elements coupling the dot to the leads. (ii) A
rather particular geometry or potential are considered. (iii)
Restrictive ranges of parameters are assumed. 

Motivated by the experiment we present here a mechanism which
accounts for phase correlations 
for different values of $V_g$.
Our theory contains two desirable features which were
missing from previous works: (i) Our mechanism is generic and does not
invoke the peculiarities listed above. (ii) We identify a large
dimensionless parameter in the light of which our theory is formulated.
Our analysis pertains to individual, disorder specific systems. In
addition we also calculate the disorder-averaged phase-phase
correlation function which depends on the gate voltage and observe 
interaction-induced correlations.
 
We argue below that as $V_g$ is varied, there are three distinct
mechanisms for $t$ to acquire a phase change of $\pi$: there is an
increase by $\pi$ as the gate voltage is swept through a resonance;
between resonances we may encounter a near-resonance phase lapse
(NRPL) and a valley phase lapse (VPL), each involving a phase change
of $\pi$.

Our sum rule states that the {\it number} of $\pi$-changes between
consecutive valleys due to all these mechanisms is even, resulting in
$\Delta \alpha=0 (\mbox{mod} 2\pi)$. The frequency of deviations from
this sum rule is small in $\Delta/U$ where $\Delta$ is the mean single
particle level spacing and $U$ is the charging energy of the dot.

We consider an Aharonov-Bohm (AB) interferometer where a QD is
embedded in one arm. The arm containing the QD can be modeled by the 
Hamiltonian
\begin{eqnarray}
H &=& H^L+H^R+H^{QD}+H^T \ , \\
H^{L(R)} &=& \sum_{k} \varepsilon_k a_k ^{L(R) \dagger} a_k ^{L(R)}\ ,
\\
H^{T} &=& \sum_{k,j} V_{j,k} ^L c_j ^{\dagger} a_k ^L + h.c. +
L \leftrightarrow R \ , \\
H^{QD} &=& \sum_{j} (\epsilon_j-e V_g) c_j ^{\dagger} c_j +\frac{U}{2}
\hat{N}(\hat{N}-1). 
\end{eqnarray}
$H^{L,R}$ describe the regions to the left and right of the
QD, $H^T$ represents the tunneling of electrons in and out of the QD,
and $H^{QD}$ describes the states of the isolated QD with the constant
interaction term.

The total transmission probability $T(E)$ through the AB
interferometer, T(E), is the
modulus squared of the sum of the transmission amplitudes 
through the two arms, $t(E)$ and $t_0$ (the latter refers to the free
arm and is assumed to be constant, $|t_0|\gg |t(E)|$).
Since the entire interferometer is coupled to external reservoirs, 
$T(E)$ needs to be convoluted with the Fermi function
$f$:
\begin{eqnarray}
T&=&\int dE \left (-\frac{\partial f}{\partial E} \right) |t(E)+t_0|^2
\simeq \nonumber\\
& &|t_0|^2 + 2 \mbox{Re} t_0 ^* \int dE \left (-\frac{\partial
  f}{\partial E} \right) t(E) \ .
\end{eqnarray}

\begin{figure}[htb]
  \begin{center}
    \leavevmode
    \parbox{0.45\textwidth}
{\psfig{file=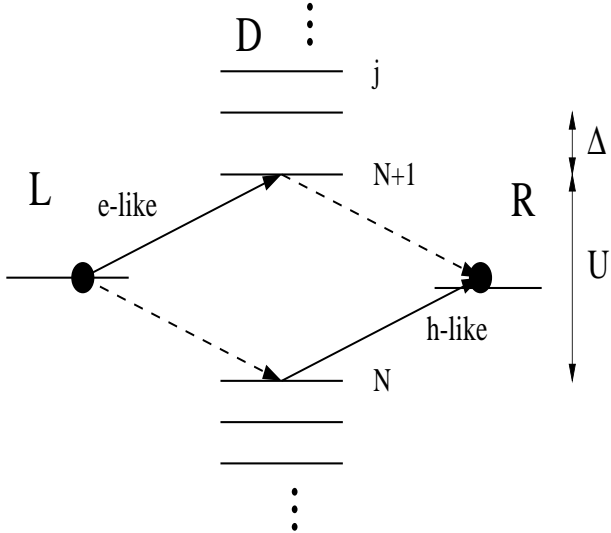,width=0.45\textwidth,height=0.4\textwidth,angle=0}}
  \end{center}
  \caption[]{
    Level scheme out of resonance, depicting electron-like and
    hole-like processes.}
  \label{sketch}
\end{figure}

We first propose a qualitative picture which motivates and expounds
the phase correlations. Fig. \ref{sketch} depicts (schematically)
virtual processes (second order in the $V$'s) which, at zero
temperature, contribute to the transmission amplitude, say, from the
left lead (L) to the right lead (R). An off-resonance (valley) set-up 
is shown.
There are electron-like processes, employing vacant levels
(e.g. $\epsilon_j$) as intermediate states. The contribution of such a
process to $t(E)$ is (neglecting corrections due to level
broadening) $\sim V_j^{L}V_j^{R*}/
(E-(\epsilon_j-eV_g+U\cdot N)+i\Gamma)$ where
$V_j=V_{j,k(E)} \sqrt{2\pi \rho(E)}$ with $N$ electrons on the QD. 
Here, $\rho$ is the density of
states in the left or right lead. We focus on a disordered QD where
$\{ \epsilon_j \}$ and $\{ V_j \}$ are fluctuating and are described
to a good approximation by Random Matrix Theory. The numerator thus
has a random sign (in the absence of a magnetic field it can be chosen
real). The contribution of the electron-like processes to $t(E)$ arise
from a large number of random terms (of which $\sim U/\Delta$
contribute significantly). We denote the sum over the contributions
with $j\ge N+2$  ($j\le N-1$) the ``electron team'' (``hole team'')
\cite{terminology}. The contribution of the $j=N+1$ level ($j=N$) will
be referred to as the ``electron team captain'' (``hole team 
captain'').

The following observations are now due: (a) The signs of the four
contributions (the ``teams'' and the ``captains'') are random; however,
the teams in the $N^{th}$ valley and in the $(N+1)^{th}$ valley differ
by very little from each other (essentially by the contribution of one
level). Thus, up to events which are rare by the parameter $\Delta/U$,
the sign of the e-teams (h-teams) in two consecutive valleys are the
same. (b) As $e V_g$ is increased in the valley the magnitude of the
e-team (h-team) increases (decreases). Furthermore, as the resonance
is approached, the relative importance of the team diminishes and
eventually near the resonance, it is a single level -- the team
captain -- which governs the transmission.

\begin{figure}[htb]
  \begin{center}
    \leavevmode
    \parbox{0.45\textwidth}
{\psfig{file=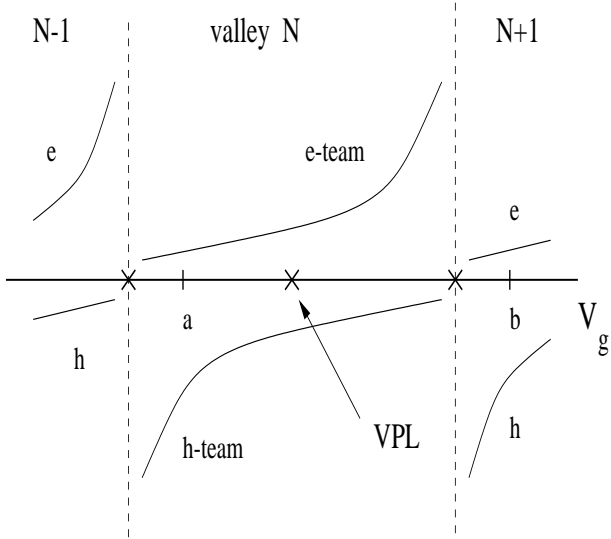,width=0.45\textwidth,height=0.4\textwidth,angle=0}}
  \end{center}
  \caption[]{
    The evolution of the e-contribution and the h-contribution to Re
    $t$
    as function of $V_g$ (schematic); $\times$ denotes a phase change
    of $\pi$. See text for the choice of the signs for the ``teams''
    and ``captains'' contributions. The total phase change over a
    period (say from $V_g(a)$ to $V_g(b)$) is $0$ (mod$\pi$). }
  \label{teams}
\end{figure}

A phase lapse occurs when the signs of the e-team and the h-team
differ (VPL) or when a team does not agree in sign with its captain
(NRPL). Fig. \ref{teams} depicts the evolution of the e-contributions
and the h-contributions to Re $t$ over a range of $V_g$. In
Fig. \ref{teams} we display the signs of the four contributions to be:
e-team$=+$, e-team captain$=+$, h-team$=-$, h-team captain$=-$. A VPL occurs
and the total number of phase gains by $\pi$ over a period is 2,
rendering $\Delta \alpha=0 (\mbox{mod} 2\pi)$. Evidently one needs to
examine each of the 16 possible sign assignments, each yielding a
different pattern of $t$ as function of $V_g$. But remarkably enough
we find that over a ``period'' as defined above
$\pi \cdot$ (at resonance)+$\pi \cdot $ (number of NRPL)+$\pi \cdot $
(number of VPL) =even.

Next we put the above picture in a more quantitative framework. The
transmission amplitude through the interacting system is linked to
the retarded Green's function $G_{ij}$ of the QD coupled to the leads
by $t(E)=\sum_{ij} V^L_i V^{R*}_j G_{ij}$. 
As we are interested in the
elastic cotunneling contributions \cite{avnaz} rendering $G_{ij}$
diagonal (tunneling in and out of the same dot state), $G_{ij}$
can approximately be determined by iterating the equation of motion. 
Specifically we use an extension of Ref. \cite{yy} to many levels in
the dot,
\begin{eqnarray}
t(E)&=& \sum_j V^L_j V^{R*}_j \sum_{N=0}^{\infty} P_N \left [
\frac{\langle n_j \rangle_N}{E-(\epsilon_j -\epsilon_N - x U)
  +i\Gamma_j} \right. \nonumber\\
&+&\left. \frac{1-\langle n_j \rangle_N}{E-(\epsilon_j-\epsilon_N
+ (1-x) U) +i\Gamma_j} \right ].
\label{trans}
\end{eqnarray} 
$\langle ... \rangle_N =\mbox{tr}_N \exp -\beta H^{QD}
.../\mbox{tr}_N \exp -\beta H^{QD}$ denotes the thermal average with
$N$ electrons. The probability to find $N$ electrons on the QD is
given 
by $P_N=\langle \hat{N} \rangle_N/\sum_M \langle \hat{N}
\rangle_M$. Deep in the valley $N$ we have $P_M\simeq\delta_{M,N}$.
The two terms of (\ref{trans}) describe the h-like and the
e-like contributions, respectively \cite{caveat}.

For the sake of simplicity the statistics we first introduce is a toy
model (to be revoked later), with the assumptions:
(i) The level spacing is constant $\epsilon_j=j
\Delta$ mimicking the level repulsion and (ii) $V_j^L V_j^{R *}=V
\eta_j$ where $\eta_j$ is a random variable which can take the values
$+1$ and $-1$ with equal probability. This models the fluctuations in
the wave function due to disorder. The relevant physics is
already contained in the toy model as a comparison with RM-generated
energies and couplings reveals. Let us first consider 
the non-interacting case. The transmission amplitude then becomes 
\begin{eqnarray}
t(E)&=&V \sum_j \frac{\eta_j}{E-(\epsilon_j- e V_g) +i\Gamma} \ ,
\end{eqnarray} 
where the system can be tuned in/out of resonance by the gate voltage 
$e V_g$. At the resonance $\alpha$ increases by 
$\pi$. The signs of $\eta_j$ govern the 
phase evolution between the resonances. If $\eta_1 \cdot \eta_2>0$
there is a decrease by $\pi$ (phase lapse) between the resonances 1 and 2, 
for $\eta_1 \cdot \eta_2 <0$ this phase lapse is absent. Note that
in a one-dimensional symmetric potential $\eta_j$ alternates in sign
implying no phase lapse. For a disordered non-interacting QD the phase
lapses occur at random \cite{phasenicorr}. Here we show that 
interaction changes this picture considerably.

We note that near the resonances $(N-1,N)$ the main contribution to
$t$ in (\ref{trans}) comes from the electron contributions of the $N-1$
valley and the hole contribution of the $N$
valley. The other contributions are
smaller by a factor of $U$. At the resonance $(N-1,N)$ the level
$\epsilon_N$ is resonating which is the e-team captain in the $N-1$ valley and
the h-team captain in the $N$ valley. When we single out the state 
$\epsilon_N$ in the
electron and hole states we see that the vicinity of $(N-1,N)$   
is equivalent to the single particle resonance of level
$\epsilon_N$, $\eta_N/(E-(\epsilon_N-e V_g)+i\Gamma)$ and background
term $A_{N-1}$ and $B_N$ corresponding to the e-team ($N-1$ valley) 
and h-team ($N$ valley). These background terms for different valleys are
strongly correlated, e.g. the level $\epsilon_{N+1}$ contributes both
to $A_{N-1}$ and $B_{N+2}$. 
\begin{figure}[htb]
  \begin{center}
    \leavevmode
    \parbox{0.45\textwidth}
{\psfig{file=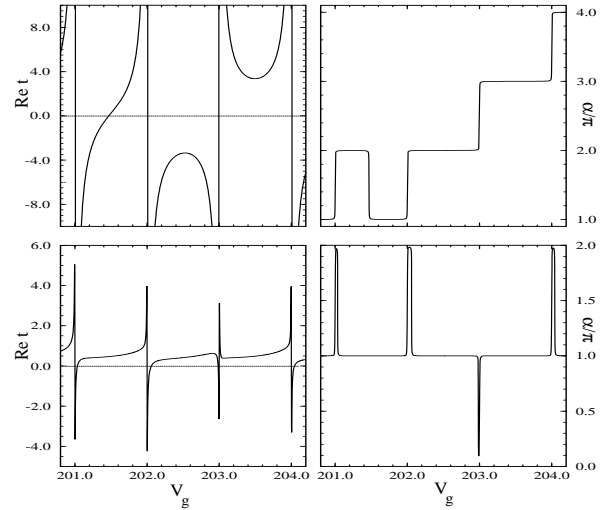,width=0.45\textwidth,height=0.4\textwidth,angle=0}}
  \end{center}
  \caption[]{Re $t$ (left) and phase (right) for a specific sequence of
    resonances, for $U=0$ (upper panel) and $U=60\Delta$ (lower
    panel).}
  \label{phaseevolu}
\end{figure}
Fig. \ref{phaseevolu} shows the evolution of Re $t$ and $\alpha$ for a
specific series of couplings for $U=0$ (upper panel) and $U=60\Delta$
(lower panel) and $kT=\Delta/12$. We focus on four resonances (at
integer values of $V_g$) with
$\eta_{1}=\eta_{2}=-\eta_{3}=\eta_{4}=1$. For $U=0$ we have a
phase lapse in the valley between the first and second resonance 
(Re $t$ becomes
zero there) and no phase lapses between the others, as explained
above. For interacting electrons we observe near resonance phase
lapses (NRPLs) due to the
background terms which make the phase
stay at $\alpha=\pi$ for almost all values of $V_g$ -- except near the
resonances. 

\begin{figure}[htb]
  \begin{center}
    \leavevmode
    \parbox{0.45\textwidth}
{\psfig{file=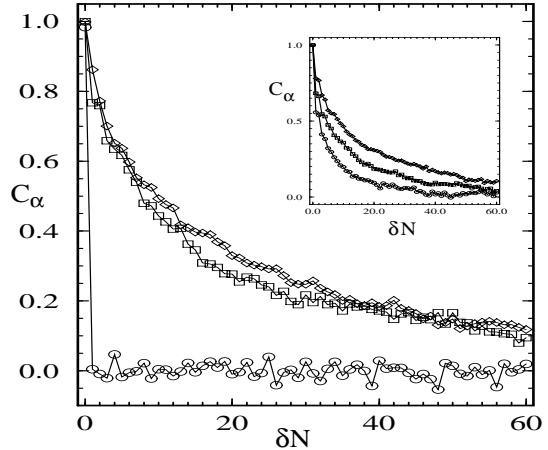,width=0.45\textwidth,height=0.4\textwidth,angle=0}}
  \end{center}
  \caption[]{Correlation function $C_{\alpha}$ vs. valley number for
    $kT=\Delta/12$. Interacting electrons (boxes and
    diamonds, $U=50\Delta$, $x_1=x_2=0.5$) show strong correlations, 
    non-interacting
    electrons (circles, $x_1=x_2=0.25$) do not. Toy model statistics
    (boxes) agrees
    well with Random Matrix model statistics (diamonds).
    Inset: Increasing the interaction
    increases the correlations (upper curve: $U=50\Delta$, middle:
    $U=25\Delta$, lower: $U=12\Delta$).}
  \label{correl}
\end{figure}

For the impurity-averaged correlations we define
$C_t=\langle t(x,N) t^*(\bar{x}, N+\delta N) \rangle / \sqrt{\langle
  |t(x,N)|^2 \rangle \langle |t(\bar{x}, N+\delta N)|^2 \rangle}$ and  
$C_{\alpha}=\langle \cos
\alpha(x,N) \cos \alpha(\bar{x},N+\delta N) \rangle$.
The calculation  of  $C_t$ follows \cite{aleinerglaz}. The
transmission amplitude (Eq. \ref{trans}) is given by 
 $t(E)\sim V^L V^R \int 
  d\omega \left[ G_{\omega}^A(L,R)-G_{\omega}^R(L,R)
 \right] G_{\omega}^{\rm ret} (E,N,x)$ where $V^{L,R}$ describe the
 left and right barrier.  
 We note that this expression consists of a disorder-dependent (but
 interaction-independent) factor, and one, $G^{\rm ret}$,
  which includes the interaction (but not the disorder).
 One then readily obtains
\begin{eqnarray}
C_t (x,N,x,N+\delta N)&=& u \frac{x(1-x)}{\delta N} [\log
\left(1+\frac{\delta N}{x u} \right) +\nonumber \\
& & \log \left(1+\frac{\delta
    N}{(1-x)u} \right ) ]
\label{ct}
\end{eqnarray}
with $u=U/\Delta$.
$C_t$ decays slowly on a scale of $\delta N \sim U/\Delta$. We also
observe that (i) for the non-interacting case $C_t$ falls abruptly to
zero for $\delta N=1$. (ii) The results for the toy model and for a
more realistic model, where $\epsilon_j$ and $V_j^LV_j^{R*}$ are
obtained from diagonalizing random matrices, agree well, implying that
the correlations are fairly insensitive to the way randomness
enters. These two remarks apply for $C_{\alpha}$ which is calculated
numerically.  
Fig. \ref{correl} shows $C_{\alpha}$ vs. the distance in valleys
$\delta N$ 
for $kT=\Delta/12$. Non-interacting electrons forget about their phase
already after one valley. In contrast, for interacting electrons we
observe a slow decrease of $C_{\alpha}$, showing that information about the
phase in valley $N$ is transferred to valley $N+\delta N$ \cite{fourier}. 
The inset of Fig. \ref{correl} shows the decay of $C_{\alpha}$ for different
values of the interaction. The decay is slower for stronger
interaction. 

To summarize we have proposed here a generic mechanism which gives
rise to strong transmission phase correlations. Our approximate sum
rule is subjected to errors which occur at a frequency $\sim
U/\Delta$. Our mechanism involving a large number of small random
contributions is conceptually different from recent models
\cite{yy,bhgw} which have utilized particularly strongly coupled
levels and which depend on rather specific geometric arrangements.

Comparing our analysis to experiments \cite{schuster} we note that in
the latter $\Delta>\Gamma>kT$ implying that the resolution near the
resonance may not be sufficient to observe  NRPL directly
\cite{kondotemp}. A crucial test of our theory would be to go to small
dots with small $U/\Delta$, or, even better \cite{pcmarcus} use other
gates to scramble the dot as we sweep from one valley to another,
suppressing correlations among valleys. This should lead to a
breakdown of our sign sum rule.

We are indebted to H. A. Weidenm\"uller for his participation in
various stages of this work. We acknowledge useful discussions with
E. Buks, M. Heiblum and C. Marcus. The work was partially supported by
 the German-Israeli Foundation , by  DIP  and by
  the Minerva Foundation. Work at the Weizmann
Institute was supported by the Center of Excellence of the Israeli
Academy of Science and Humanities.


\begin{references}
\bibitem{caveat}
  {We note that (\protect{\ref{trans}}) is an approximation near the 
   resonances. In order to describe low temperature properties 
   like the Kondo effect higher iterations of the equation
   of motion would be necessary \cite{itera}. 
   However, since we focus on the resonant
   tunneling regime $\hbar \Gamma\lesssim kT \le \Delta$ we assume that the
   resonances have a width given by the unrenormalized $\Gamma_i=2\pi
   \rho (|V_i ^L|^2 +|V_i ^R|^2)$ \cite{koenig}.}
\bibitem{fourier}
  {As an alternative one may calculate the Fourier transform with
  respect to $\delta N$ of $\cos \alpha (N,x) \cos \alpha (N+\delta
  N,x)$. This quantity underlines genuine long-range correlations,
  eliminating ``noise'' due to the random locations of zeroes within a
  valley.}
\bibitem{kondotemp}
   {At lower $T$ Kondo physics might be relevant.}
\end{references}
\end{document}